\documentclass{kapproc} 
\let\footnote\savefootnote
\let\footnotetext\savefootnotetext 
\let\footnoterule\savefootnoterule 
\kluwerbib

\RequirePackage{graphicx}%
\RequirePackage{epsf}%

\begin{document}

\articletitle{Strong MgII Absorption Systems in QSOs from the Sloan Digital
Sky Survey Early Data Release.}

\author{Daniel B. Nestor, Sandhya Rao, David Turnshek, Eric Furst\thanks{also Bucknell University}}

\affil{University of Pittsburgh}
\email{dbn@phyast.pitt.edu}

\chaptitlerunninghead{Metal Absorption in SDSS EDR QSOs.}

\section{Introduction}
The Sloan Digital Sky Survey (SDSS) is obtaining
multicolor images over more than 10,000 square degrees of high 
Galactic latitude sky and providing medium resolution spectra 
for approximately 10$^6$ galaxies and 100,000 quasars.  The early
data release (EDR) of June 2001 (Schneider et al. 2002) 
contains spectra of $\approx 3800$
QSOs with redshifts ranging from $z=0.15$ to $z=5.03$. The 
spectra cover the wavelength interval
3800\AA\ $< \lambda <$ 9200\AA\ and have resolutions ranging from 
1800 to 2100.

In order to study intervening 
low-ionization metal absorption-line systems, we are constructing
samples of systems selected for the strength of their 
$\lambda$2796 MgII lines from the EDR.  
There are 640 absorption systems in our unbiased sample 
with $\lambda$2796 REW$> 1.0$ \AA\ detected at 5$\sigma$ and with 
redshifts  $0.37 < z < 2.27$. This sample is an order of magnitude
larger than previous similarly selected samples. In this 
contribution, we present preliminary results on the statistical
properties of the sample. 


\begin{figure}
\plottwo{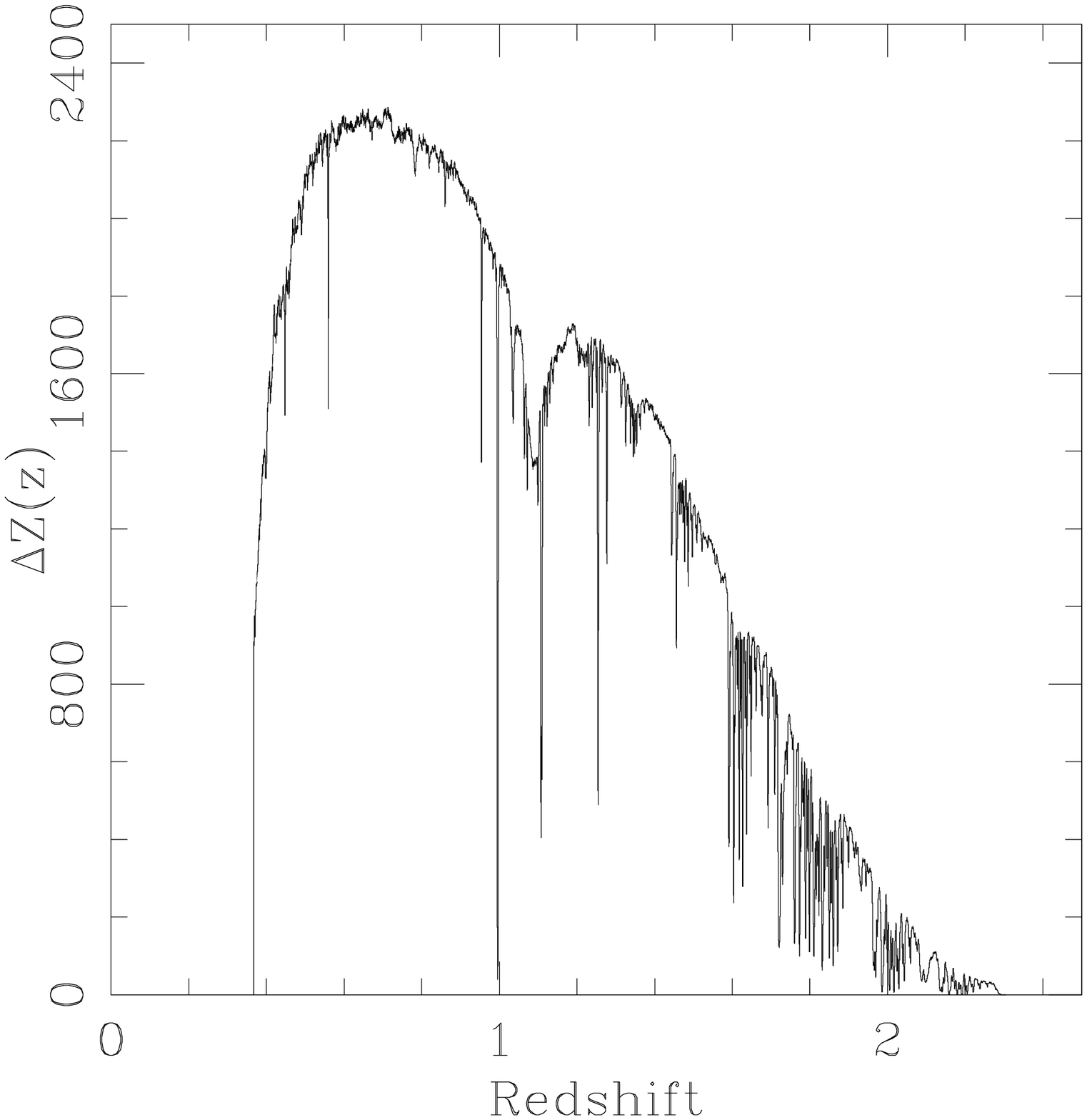}{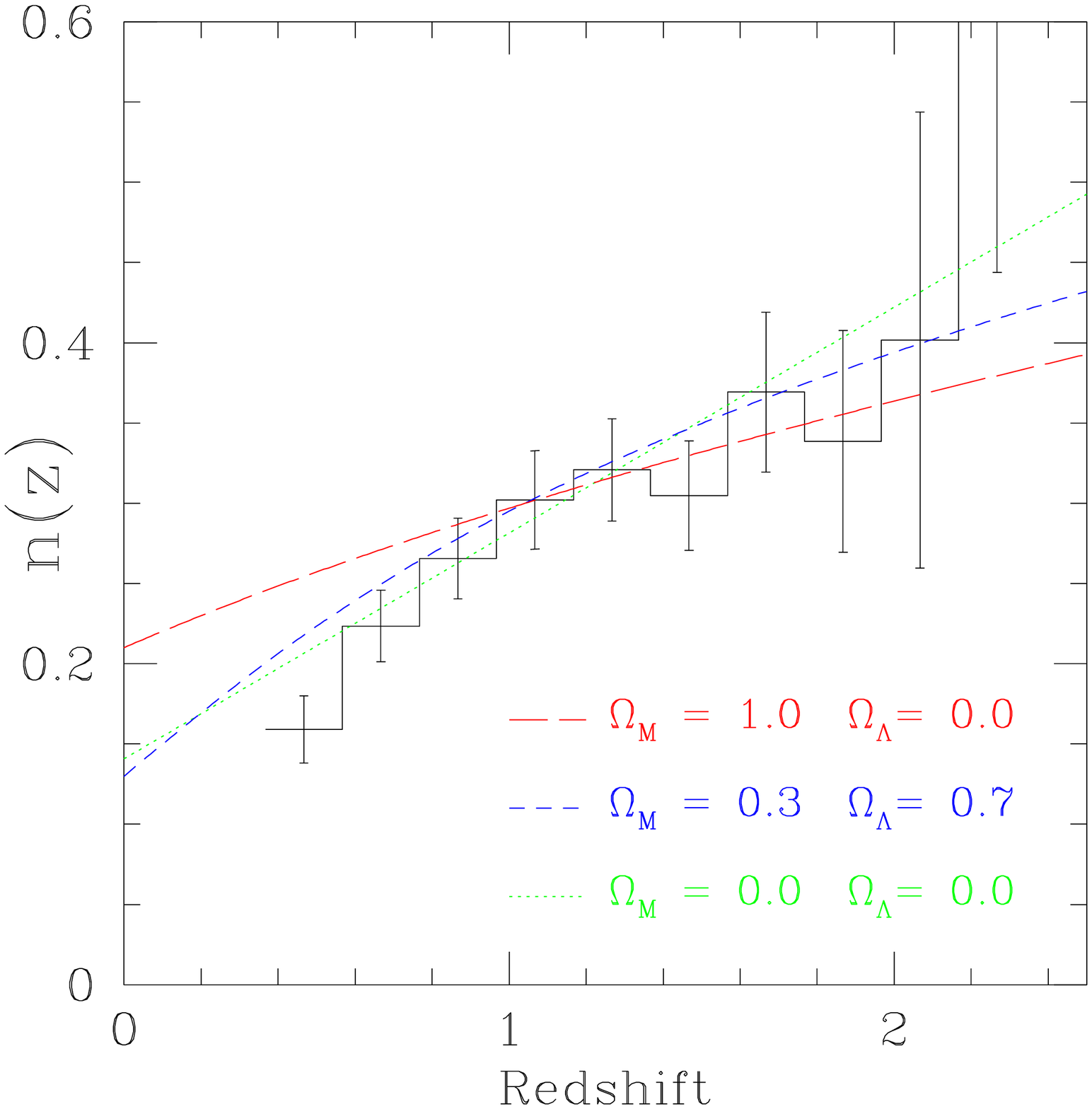}
\caption{{\bf Left:} The redshift path covered by the sample.
{\bf Right:} The number density distribution with redshift 
of the sample. Also shown
are no-evolution predictions for three cosmologies
 normalized to minimize the $\chi^2$ fit to the binned data.
The distribution is consistent with no evolution except at 
the lowest redshifts.}
\end{figure}

\begin{figure}
\plottwo{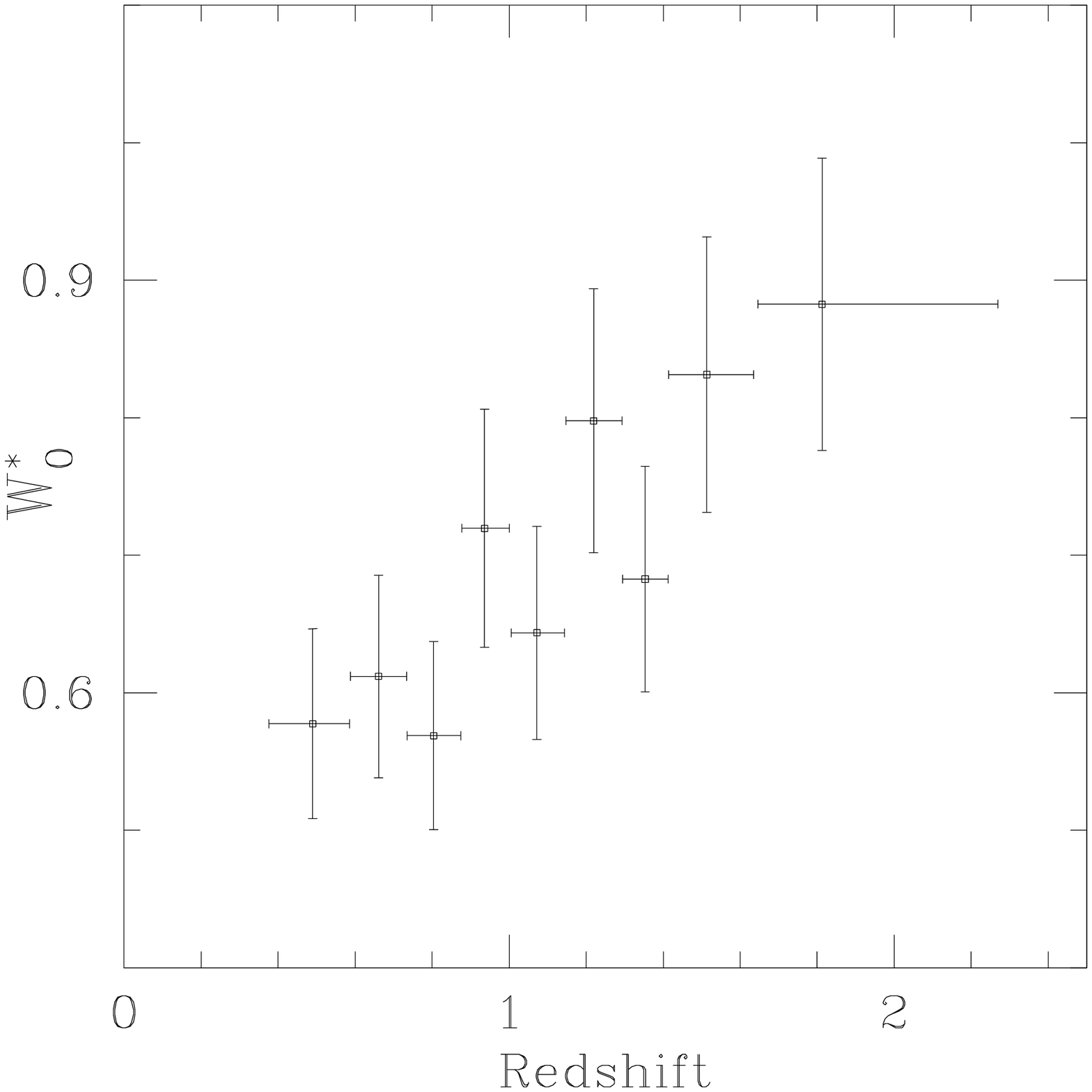}{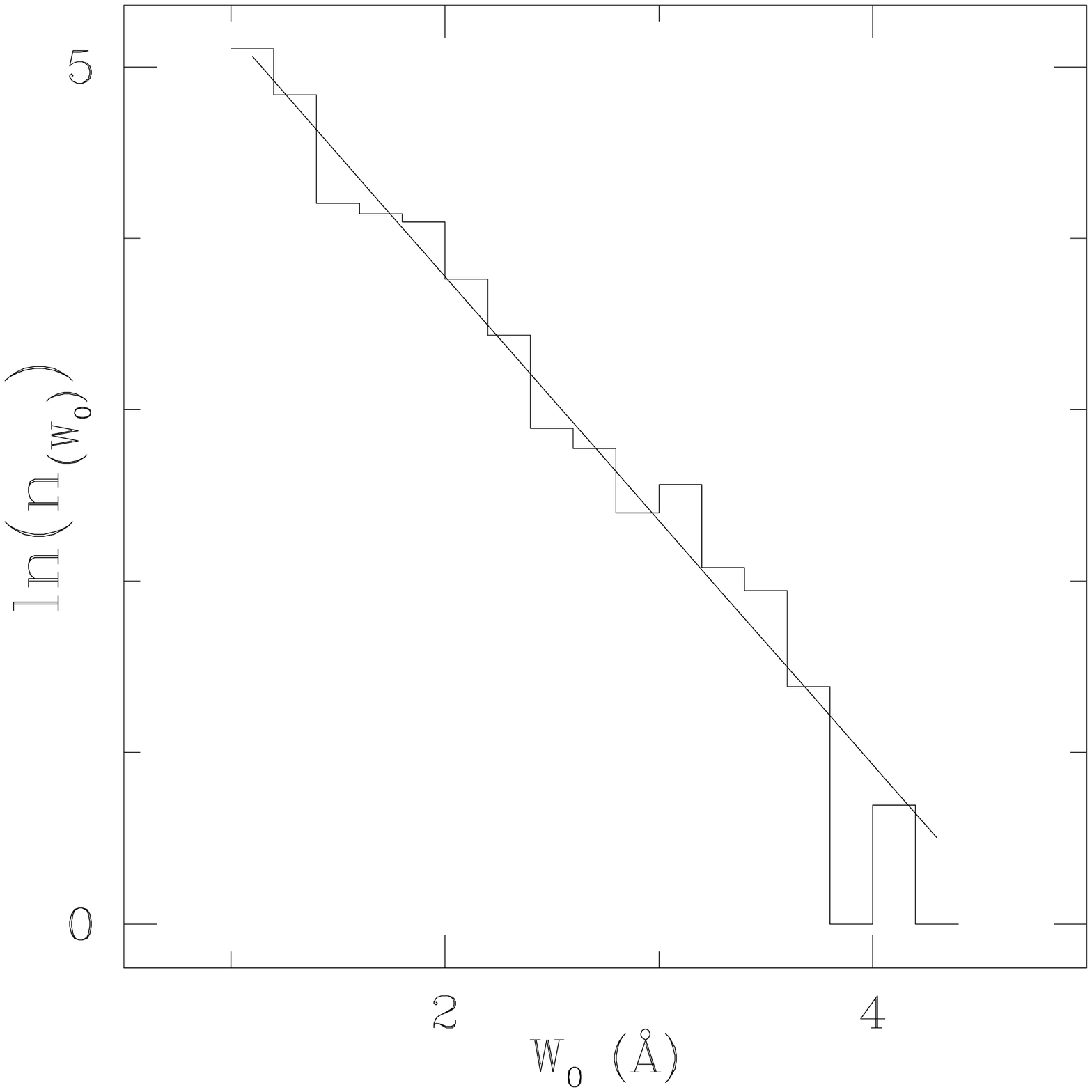}
\caption{{\bf Left:} Evolution of the REW distribution slope with 
redshift for $n(W_0)=n_0 e^{-W_0/W_0^*}$.  The data are shown divided
into redshift bins indicated by horizontal bars
with an equal number of lines in each bin.
Vertical bars are 1$\sigma$ errors.
There is mild evolution in the distribution of REWs.
{\bf Right:}
The REW distribution for the total sample.  The slope of the
maximum-likelihood fit is $-1.4$, corresponding to $W_0^*=0.7$.} 
\end{figure}

\end{document}